\documentclass[twocolumn,pra,aps,twocolumn]{revtex4}
\usepackage[latin9]{inputenc}
\usepackage{amsbsy}
\usepackage{amstext}
\usepackage[unicode=true,pdfusetitle,
 bookmarks=false,
 breaklinks=false,pdfborder={0 0 1},backref=false,colorlinks=false]
 {hyperref}
\hypersetup{
 bookmarksnumbered=false,bookmarksopen=false}

\makeatletter

\@ifundefined{textcolor}{}
{%
 \definecolor{BLACK}{gray}{0}
 \definecolor{WHITE}{gray}{1}
 \definecolor{RED}{rgb}{1,0,0}
 \definecolor{GREEN}{rgb}{0,1,0}
 \definecolor{BLUE}{rgb}{0,0,1}
 \definecolor{CYAN}{cmyk}{1,0,0,0}
 \definecolor{MAGENTA}{cmyk}{0,1,0,0}
 \definecolor{YELLOW}{cmyk}{0,0,1,0}
}

\usepackage{amsmath}
\usepackage{graphicx}
\usepackage{amssymb}
\usepackage{txfonts,color}
\makeatother

\begin{document}

\title{Arbitrary Nuclear Spin Gates in Diamond Mediated by a NV-center Electron Spin}
\author{J. Casanova}
\affiliation{Institute of  Theoretical Physics and IQST, Albert-Einstein-Allee 11, Universit\"at
Ulm, D-89069 Ulm, Germany}
\author{Z.-Y. Wang}
\affiliation{Institute of  Theoretical Physics and IQST, Albert-Einstein-Allee 11, Universit\"at
Ulm, D-89069 Ulm, Germany}
\author{M. B. Plenio}
\affiliation{Institute of  Theoretical Physics and IQST, Albert-Einstein-Allee 11, Universit\"at
Ulm, D-89069 Ulm, Germany}

\begin{abstract}
We show that arbitrary $N$-qubit interactions among nuclear spins can be achieved efficiently 
in solid state quantum platforms, such as nitrogen vacancy centers in diamond, by exerting control only 
on the electron spin  coupled to the nuclei. This allows to exploit nuclear spins as robust quantum 
registers and the direct measurement of nuclear many-body correlators. The method takes advantage of 
recently introduced dynamical decoupling techniques and avoids the necessity of external, slow, control 
on the nuclei. Our protocol is general, being applicable to other nuclear spin based platforms with 
electronic spin defects acting as mediators  as silicon carbide.
\end{abstract}
\maketitle

\section{Introduction} Nuclear spins in solid-state platforms such as diamond or silicon
carbide are natural and reliable quantum registers with exceptional coherence times~\cite{Maurer12}.
In order to realise the potential of this resource for quantum computing~\cite{Nielsen},
quantum simulations~\cite{Georgescu14}, and quantum sensing~\cite{Degen16, Wu16}, it is necessary
to achieve both individual nuclear spin rotations and arbitrary coherent coupling between nuclear
spins. In close proximity, nuclear spins exhibit a natural coupling because of their intrinsic
nuclear dipolar interactions which may be exploited for quantum simulations \cite{Cai13}. However,
this natural coupling is generally small as a consequence of the weak nuclear magnetic moments
\cite{Georgescu14}, while the available closest internuclear distances are always lower bounded by
the lattice constants of the host material. Furthermore, the spin active nuclear isotopes
appear in the platforms of interest with a low natural abundance, e.g. $1.1\%$ and $4.7\%$  for
the cases of $^{13}$C and $^{29}$Si nuclei which are the relevant species for diamond~\cite{Doherty13, Dobrovitski13} and silicon carbide~\cite{Baranov05, Seo16}. Furthermore, the positions of the nuclei are fixed which make the modulation of the internuclear coupling challenging.

The  tuning of the direct coupling of nuclear spins is excessively demanding, although
it can be suppressed (up to a certain degree) by the application of suitably tuned radio
frequency (rf) fields~\cite{Lee65, Michal08,Cai13}. However, nuclear spins can be coupled to external
magnetic fields through Zeeman interactions~\cite{Abragam61}, and more importantly, they can couple
to nearby electron spins. In this respect, the electron spin of a NV center is a promising nano-scale
device to detect and control nuclear spins using the electron-nuclear hyperfine interaction~\cite{Doherty13, Dobrovitski13}. The large magnetic moment of electron spins allows a single NV center to couple
strongly to many nuclear spins. In addition, with dynamical control achieved due to the application on
the NV center of decoupling sequences such as Carr-Purcell-Meiboom-Gill (CPMG)~\cite{Carr54, Meiboom58},
pulse arrangements of the XY kind~\cite{Maudsley86, Gullion90}, and adaptive XY sequences
\cite{Casanova15, Wang16, Casanova16}, one can entangle the NV electron to individual nuclear spins
(including the $^{14}$N nucleus inherent to the NV center) for the sake of electron-nuclear two-qubit
quantum gates and quantum algorithms~\cite{Gurudev07, Neuman10, Robledo11, vanderSar12, Kolkowitz12, Taminiau12, zhao12, Liu13, Taminiau14, Waldherr2014, Cramer16, Hanson2015}.

Particularly, the set of techniques developed in~\cite{Casanova15, Wang16, Casanova16} allows
for highly selective and robust electron-nuclear quantum gates evolving according to the
Hamiltonians $\sigma_z I_x$ or $\sigma_z I_y$ by using sequences of non-equidistant microwave decoupling
pulses. Note that $\sigma_z = |m_s = \pm 1 \rangle\langle m_s \pm 1 | - |m_s = 0  \rangle\langle m_s = 0 |$ corresponds to an effective electronic spin-$\frac{1}{2}$ operator~\cite{Doherty13, Dobrovitski13}, while
$I_x$, $I_y$ are nuclear spin operators with $I=\frac{1}{2}$. In addition, in~\cite{Wang16bis} it is shown
how the judicious application of a delay window achieves an interaction of the kind $\sigma_z
I_z$. Furthermore, these techniques can incorporate a decoupling rf field \cite{Wang16, Casanova16, Wang16bis}
to combine electron-nuclear entangling gate generation with a suppression of the internuclear decoupling.

In this work we show that arbitrary single, and $N$-body nuclear spin interactions can be realised
efficiently in the frame of electron spin defects with nearby nuclear spins through a specific
combination of selective electron-nuclear gates. The latter can be achieved when the natural hyperfine
couplings between the electronic- and nuclear-spins are appropriately modulated with dynamical decoupling 
techniques. Our method combines the two key advantages of electron and nuclei qubits, namely fast electronic 
control and the long nuclear spin coherence times. In addition, we demonstrate how the same techniques 
allow to measure directly nuclear many-body correlators. To exemplify the protocol we use NV centers in 
diamond as the model system, but our method is general and can be used in other solid-state quantum 
platforms such as silicon carbide.

\section{Elementary entangling gates} With the dynamical decoupling techniques in~\cite{Casanova15, Wang16, Casanova16, Wang16bis},
one can achieve highly selective entangling quantum gates of the form
\begin{equation}
    Q_{j}^{\alpha}(\varphi)=\exp\left(i\varphi\sigma_{z}I_{j}^{\alpha}\right),
\end{equation}
between the NV electron spin (with Pauli operator $\sigma_{z}$) and the $j$-th nuclear
spin (with the spin operator $I_{j}^{\alpha}$ and $\alpha=x,y,z$) while prolonging the electron spin coherence. Another important
feature is that, as opposed to standard dynamical decoupling methods~\cite{Carr54, Meiboom58, Maudsley86, Gullion90}, the phase
$\varphi$ is fully tunable. We will see later that this is a crucial requirement for
designing our quantum algorithm. In addition, with microwave control, single qubit gates
can be applied to the NV electron spin. These are, for instance, $X_{\phi}=e^{i\phi\frac{\sigma_{x}}{2}}$ i.e. a rotation of an arbitrary and controllable phase $\phi$ around the $x$ axis, although any other direction is available.

\section{Protocol for N-body nuclear interactions} The elementary gates presented before
permit the implementation of $N$-body interactions (denoted in the following by $ U_{\phi}$) between the nuclear spins that can
be individually addressed by the NV center. In the last section we will demonstrate numerically that the individual nuclear addressing is possible even when a certain target spin is surrounded by other nuclei interacting with the NV center. The latter is of great benefit when dealing with dense samples because they contain a potentially large number of available nuclear qubits.  For example, by having the entangling gates  $Q_{j_{1}}^{\alpha_{1}}(\varphi_{1})Q_{j_{2}}^{\alpha_{2}}(\varphi_{2})$ at
hand, one can demonstrate the following equality (up to an irrelevant global phase):

\begin{eqnarray}\label{Uphi}
    U_{\phi} & = & Q_{j_{1}}^{\alpha_{1}}(\varphi_{1})Q_{j_{2}}^{\alpha_{2}}(\varphi_{2})X_{2\phi+\pi}Q_{j_{1}}^{\alpha_{1}}(\varphi_{1})Q_{j_{2}}^{\alpha_{2}}(\varphi_{2})X_{\pi} \nonumber\\
    & = & e^{i \phi  \sigma_x (\cos\varphi_{1} - 2i\sin\varphi_{1}\sigma_{z}I_{j_1}^{\alpha_1})(\cos\varphi_{2} - 2i\sin\varphi_{2}\sigma_{z}I_{j_2}^{\alpha_2})},
\end{eqnarray}
which contains two-qubit, and many-body (in this specific case three-body) interactions
involving the electron and the nuclear spins, see Appendix for a detailed derivation of Eq.~(\ref{Uphi}).

We want to note that an especially interesting situation is realised when $\varphi_{1}
=\varphi_{2}=\frac{\pi}{2}$, i.e. when we are making interact with the same phase  the electron spin with different nuclei. In this case we have $U_{\phi}=\exp\left[- i 4 \phi\sigma_{x}I_{j_1}^{\alpha_1}I_{j_2}^{\alpha_2}\right]$ namely a three-body time evolution
operator where the phase $\phi$ corresponds to the one of the central gate, $X_{2\phi+\pi}$,
in the first line of Eq.~(\ref{Uphi}).

Now it is possible
to generalise the results in Eq.~(\ref{Uphi}) and demonstrate that with the following sequence of gates
\begin{eqnarray}\label{recipe}
    U_{\phi} & = & Q_{N}X_{2\phi+\pi}Q_{N}X_{\pi},
\end{eqnarray}
where $Q_{N}=\prod_{n=1}^{N}Q_{j_{n}}^{\alpha_{n}}(\varphi_{n})$, $N$ labels the total number
of addressable nuclear spins,  $\alpha_{n}=x,y,z$, and by noting that $[Q_{j_{n}}^{\alpha_{n}}(\varphi_{n}), Q_{j_{m}}^{\alpha_{m}}(\varphi_{m}) ] = 0$ for $n \neq m$, i.e. the different entangling gates
commute, one can find that, for $\varphi_{n}=\frac{\pi}{2}\ \forall n$, and in the case $N=2M$
(i.e. we are subsequently addressing an even number of nuclear spins) the time evolution operator is
\begin{eqnarray}\label{generaleven}
    U^e_{\phi} & = & \exp\left[(-1)^M i \ 2^{2M}\phi \ \sigma_{x}\prod_{n=1}^{2M}I_{j_{n}}^{\alpha_{n}}\right].
\end{eqnarray}
In the same manner,  for $N=2M + 1$ (an  odd number of nuclear spins are addressed) we have
\begin{eqnarray}\label{generalodd}
    U^o_{\phi} & = & \exp\left[(-1)^{(M+1)} i \ 2^{2M+1}\phi \ \sigma_{y}\prod_{n=1}^{2M+1}I_{j_{n}}^{\alpha_{n}}\right].
\end{eqnarray}
Note that in both cases we have a $(N+1)$-body evolution that involves the electron and nuclear spins.

Now, if the $U^e_{\phi}$ or $U^o_{\phi}$ operators act on an initial state such that, for
the even case, we have $\rho^e_{t=0} = |x_{\pm}\rangle \langle x_{\pm}| \otimes \rho_N$
where $\sigma_x |x_{\pm}\rangle = \pm |x_{\pm}\rangle$, while the initial state for the odd case is
$\rho^o_{t=0} = |y_{\pm}\rangle \langle y_{\pm}| \otimes \rho_N$ with $\sigma_y |y_{\pm}
\rangle = \pm |y_{\pm}\rangle$ and  $\rho_N$ represents an initial nuclear state, we have
the following two possibilities
\begin{equation}
    U^e_{\phi} \ \rho^e_{t=0} (U^e_{\phi})^\dag = |x_{\pm}\rangle \langle x_{\pm}|  \otimes \tilde{U}^e_{\pm\phi} \   \rho_N \ (\tilde{U}^e_{\pm\phi})^\dag
\end{equation}
and
\begin{equation}
    U^o_{\phi} \ \rho^o_{t=0} (U^o_{\phi})^\dag = |y_{\pm}\rangle \langle y_{\pm}|  \otimes \tilde{U}^o_{\pm\phi} \   \rho_N \ (\tilde{U}^o_{\pm\phi})^\dag,
\end{equation}
where the $N$-body nuclear operators $\tilde{U}^e_{\pm\phi}$, $\tilde{U}^o_{\pm\phi}$ read
\begin{equation}\label{even}
\tilde{U}^e_{\pm\phi} = \exp\left[-(1)^M i \ 2^{2M} (\pm\phi) \prod_{n=1}^{2M}I_{j_{n}}^{\alpha_{n}}\right],
\end{equation}
and
\begin{equation}\label{odd}
\tilde{U}^o_{\pm\phi} = \exp\left[-(1)^{(M+1)} i \ 2^{2M+1}(\pm\phi) \prod_{n=1}^{2M+1}I_{j_{n}}^{\alpha_{n}}\right].
\end{equation}
It is also important to note that the gates $X_{2\phi+\pi}$ and  $X_{\pi}$ in Eq.~(\ref{recipe})
can be replaced by $Y_{2\phi+\pi}$ and $Y_{\pi}$ where $Y_{\phi} = e^{i\phi \frac{\sigma_y}{2}}$,
or for any other gate that implies a rotation on the XY plane. The latter would give rise to a set 
of results similar to the ones in Eqs.~(\ref{generaleven}), (\ref{generalodd}), (\ref{even}), and
(\ref{odd}). Finally, note that, during gate performance, the electron spin gets selectively 
coupled with different target nuclei of the sample but will also be affected by different 
noise sources (see later for a description of the typical error sources in the case of NV 
centers in diamond). Therefore, since the electron spin is the mediator of nuclear interactions, 
its quantum state has to be protected against errors during the protocol which we achieve by means
of dynamical decoupling techniques.

In this manner we have realised an effective $N$-body interaction acting on a set
of nuclei, while the electron spin gets uncoupled after the process. We would like to
note that Eqs.~(\ref{even}) and (\ref{odd}) allows one to couple distant nuclear spins and,
remarkably, the achieved phase $\phi$ can be easily extended without affecting significantly
to the total time of the protocol. In this respect note  that $\phi$ is introduced through
a single-qubit gate on the electron spin that can be implemented in a time on the order of
nanoseconds. Furthermore, an electron spin rotation to change $|x_{\pm}\rangle \rightarrow
|y_{\pm}\rangle $ allows us to subsequently combine the final results in Eqs.~(\ref{even}), (\ref{odd})
in order to concatenate a set of $N$-qubit quantum gates upon different nuclei.

In the same manner, and using again the NV center spin as the interaction  mediator, a single
entangling gate $Q_{j_n}^{\alpha_n}(\varphi_n)$ can be used to individually rotate any addressable
spin by an arbitrary phase. This is achieved by simply noting that
\begin{equation}
Q_{j_n}^{\alpha_n}(\varphi_n) |z_{\pm}\rangle \langle z_{\pm}| \otimes \rho_N  (Q_{j_n}^{\alpha_n})^\dag =  |z_{\pm}\rangle \langle z_{\pm}| \otimes e^{\pm\left(i\varphi_n I_{j_n}^{\alpha_n}\right)} \rho_N e^{\mp\left(i\varphi_n I_{j_n}^{\alpha_n}\right)},
\end{equation}
where $\sigma_z |\pm z \rangle = \pm |\pm z \rangle$. Hence, single nuclear spin rotations can be applied without having to introduce weak control rf fields that would require further calibration of the system.

In the end of the computing process a selective SWAP gate between the electron and a target  nuclear spin allows to transfer the nuclear spin quantum state to the NV center and reconstruct the nuclear spin state, or measure a nuclear spin operator, by optical readout applied on the NV center which, at low temperatures, can achieve fidelities exceeding $95\%$ \cite{Hanson2015}. In this respect, we want to note that a SWAP gate can be performed, up to a global phase, as ${\rm SWAP}_{e,j_n} = \exp{[i\frac{\pi}{2}(\sigma_z I_{j_n}^z + \sigma_x I_{j_n}^x + \sigma_y I_{j_n}^y)]} = \exp{[i\frac{\pi}{2} \sigma_z I_{j_n}^z]} \exp{[i\frac{\pi}{2} \sigma_x I_{j_n}^x]}  \exp{[i\frac{\pi}{2} \sigma_y I_{j_n}^y]}$, where each of the previous two-qubit gates can be realised with~\cite{Casanova15, Wang16, Casanova16, Wang16bis} plus additional single-qubit gates on the electron spin. Note also that a selective iSWAP gate (with ${\rm iSWAP}_{e,j_n} = \exp{[i\frac{\pi}{2}( \sigma_x I_{j_n}^x + \sigma_y I_{j_n}^y)]} = \exp{[i\frac{\pi}{2} \sigma_x I_{j_n}^x]}  \exp{[i\frac{\pi}{2} \sigma_y I_{j_n}^y]}$) is also valid to retrieve the nuclear state information.

\section{Measuring nuclear many-body correlations} The same techniques can be applied to directly
measure the expectation value of delocalised $N$-body nuclear operators. This
can be done through the next equality  (for the sake of simplicity we develop here the case for
$U^e_{\phi}$ while the odd case, i.e. the case for $U^o_{\phi}$, is similar). After a set of $N$-body
operations we have that the final state is $\rho = \rho_e\otimes\rho_n(t)$, with $\rho_e = |\alpha_{\pm}\rangle \langle \alpha_{\pm}|$ and $\alpha = x, y$ or $z$ depending on the sequence of gates we performed. Now we can take, with an electron spin flip, the state $\rho_e$ to an eigenstate of $\sigma_z$ i.e. $\rho_e \rightarrow |1\rangle\langle 1|$ with $\sigma_z|1\rangle = |1\rangle$, apply an additional
gate $U^e_{\phi}$ and measure, for example, the $\sigma_y$ electronic operator. This process leads to
\begin{eqnarray}
    \langle \sigma_y\rangle &=& {\rm Tr}\bigg[ \rho_e\otimes\rho_n(t) (U_{\phi}^e)^\dag \sigma_y  U_{\phi}^e \bigg] \nonumber\\
    &=&  {\rm Tr}\bigg[ \rho_e\otimes\rho_n(t) \  e^{\big[(-1)^M (-i) \ 2^{2M} (2\phi) \ \sigma_x \prod_{n=1}^{2M}I_{j_{n}}^{\alpha_{n}} \big]} \  \sigma_y \bigg].
\end{eqnarray}
Now, if we select the phase $\phi$ such that $(-1)^M 2\phi = (\frac{\pi}{2} + m \ 2\pi)$ with $m \in \mathbb{Z}$, we find that $\langle \sigma_y\rangle = {\rm Tr}\bigg[ \rho_e\otimes\rho_n(t) \ \sigma_z  \prod_{n=1}^{2M}\sigma_{j_{n}}^{\alpha_{n}} \bigg]$ and  one can write
\begin{equation}
\langle \sigma_y\rangle = {\rm Tr}_n\bigg[\rho_n(t) \prod_{n=1}^{2M}\sigma_{j_{n}}^{\alpha_{n}} \bigg] = \bigg\langle \prod_{n=1}^{2M}\sigma_{j_{n}}^{\alpha_{n}} \bigg\rangle,
\end{equation}
where ${\rm Tr}_n\big[ \cdot \big]$ denotes the trace over the nuclear degrees of freedom. In this manner
a highly delocalised nuclear operator gets encoded into an easy to measure electronic expectation value.

\section{Implementation with NV centers} 
In order to demonstrate the working principle of our
protocol we will consider diamond technologies, i.e. a NV center in the presence of a nuclear
spin bath, as the target system to numerically study our method. When a strong magnetic field $B_{z}$ is aligned with the NV axis, the $\hat{z}$
direction, the Hamiltonian of the coupled system in the rotating frame of the free energy
electronic spin Hamiltonian $H_0=DS_{z}^{2}-\gamma_{e}B_{z}S_{z}$ reads ($\hbar=1$)
\begin{eqnarray}\label{modelNV}
    H = A^{\parallel} S_z -\sum_{j}\gamma_{j}B_{z}I_{j}^{z} +
    S_{z}\sum_{j}\vec{A}_{j}\cdot\vec{I}_j +H_{\rm c}.
\end{eqnarray}
Here, the first term  $A^{\parallel} S_z$ with $A^{\parallel} \approx - (2\pi)\times 2.162 $ MHz,
see~\cite{MChen15}, and the spin-1 operator $S_z = |1\rangle \langle1| - |-1\rangle \langle-1|$,
corresponds to the longitudinal component of the coupling with the $^{14}$N nucleus adjacent to
the vacancy. In our simulations, instead of using the $^{14}$N nucleus as a resource qubit, we
take it as the origin of a large detuning error with magnitude $|A^{\parallel} |$ \cite{Loretz15} for demonstrating the robustness of our
protocol. However if the $^{14}$N is polarised at the beginning of the operational
process that detuning error is negligible.  The constants  $\gamma_{e} = -(2\pi)
\times 2.8024 \ \frac{\rm MHz}{\rm G}$ and $\gamma_{j} \equiv \gamma_{^{13} \rm C} =
(2\pi) \times 1.0705 \ \frac{\rm kHz}{\rm G}$ $\forall j$ represent the electronic and nuclear (in
this case for the $^{13}$C nucleus) gyromagnetic ratios.

The interaction between the NV
center and the nuclear spins is mediated by the hyperfine vector that we will consider
as dipolar like in the simulations, i.e. $\vec{A}_{j}=\frac{\mu_{0}\gamma_{e}\gamma_{j}}{4\pi|\vec{r}_{j}|^{3}}[\hat{z} - 3\frac{(\hat{z}\cdot\vec{r}_{j})\vec{r}_{j}}{|\vec{r}_{j}|^{2}}]$, where $\vec{r}_j$ is
the vector that connects the NV center and each environmental nuclei. Note also that,
because of recently developed positioning methods~\cite{Wang16} we will take $\vec{A}_j$
as known quantities. The large zero field splitting $D=(2\pi) \times 2.87$ GHz had allowed
us to neglect non-secular components in Eq.~(\ref{modelNV}). Furthermore, an external
microwave control can be introduced  in Eq.~(\ref{modelNV}) through the term $H_{\rm c} =
\Omega (|1\rangle\langle0| e^{i\vartheta} + |0\rangle\langle1| e^{-i\vartheta})$ with
$\Omega$ being  the Rabi frequency of the microwave field. In this manner, we are selecting the electronic
spin subspace $|0\rangle$, $|1\rangle$ to define our qubit.  In our numerical simulations we will additionally
introduce an error in $\Omega$ for considering realistic experimental conditions. More
specifically, if the required time for a $\pi$-flip rotation of the electron spin is
$t_{\pi} = \frac{1}{4\Omega}$, we will effectively introduce a Rabi frequency $\Omega(1 +
\epsilon)$ with $\epsilon$ the relative error that we will set as $1\%$~\cite{Cai12}. As we will see in the Appendix, we consider this error as constant because a realistic estimation for the correlation time of amplitude fluctuations for microwave fields is $\approx 1$ms~\cite{Cai12}, which is a large quantity when compared with the time to execute each individual unit of our dynamical decoupling sequence. For more details see Appendix B (numerical simulations).
Finally $\vartheta$ is a phase that, for the sake of robustness, remains constant during
each microwave pulse but changes between pulses~\cite{Casanova15, Ryan10, Souza11}, see more details in the Appendix B.

Hence, under the action of the microwave control pulses  the final  Hamiltonian is
\begin{eqnarray}\label{modelNV2}
    H = -\sum_{j} \omega_j \ \hat{\omega}_j \cdot \vec{I}_{j} + F(t) \ \sigma_{z} \sum_{j}\vec{A}_{j}\cdot\vec{I}_j ,
\end{eqnarray}
where $\sigma_z = |1\rangle\langle 1| - |0\rangle\langle 0|$, $F(t)=\pm 1$ is the modulation
function appearing because of the action of $\pi$-pulses upon the electron spin, and
$\vec{\omega}_j = \gamma_{^{13} \rm C} B_z  \hat{z} - \frac{1}{2} \vec{A}_j$ with $\omega_j = |\vec{\omega}_j |$ and
$\hat{\omega}_j = \vec{\omega}_j/|\vec{\omega}_j |$.  In this ideal description  the detuning
term $A^{\parallel} S_z$ has been eliminated because of the external microwave driving,
however our simulations will incorporate this detuning term since they are 
performed assuming the Schr\"odinger equation associated to the Hamiltonian in Eq.~(\ref{modelNV}). In addition, we will  not consider electron relaxation processes because,
at low temperatures around $4$ K, measurements of the relaxation time ($T_1$) on the order
of seconds have already been reported~\cite{Cramer16, Jarmola12} which largely exceeds
the time for performing entangling nuclear gates. It is noteworthy that, in the case of~\cite{Cramer16} NV coherence times larger than 25 ms are achieved in high purity IIa-type diamond sample at 4.2 K. Other experiments, as the one in~\cite{Gill13}, report NV $T_2$ times of $\approx 0.6$ seconds at 77 K and, again, in samples with low nitrogen concentration. In this manner, and according to the previously commented experimental results, we will not consider the NV-electron coupling because of $P_1$ centers in the diamond lattice. 

\begin{figure}[t]
\hspace{-0.40 cm}\includegraphics[width=0.82\columnwidth]{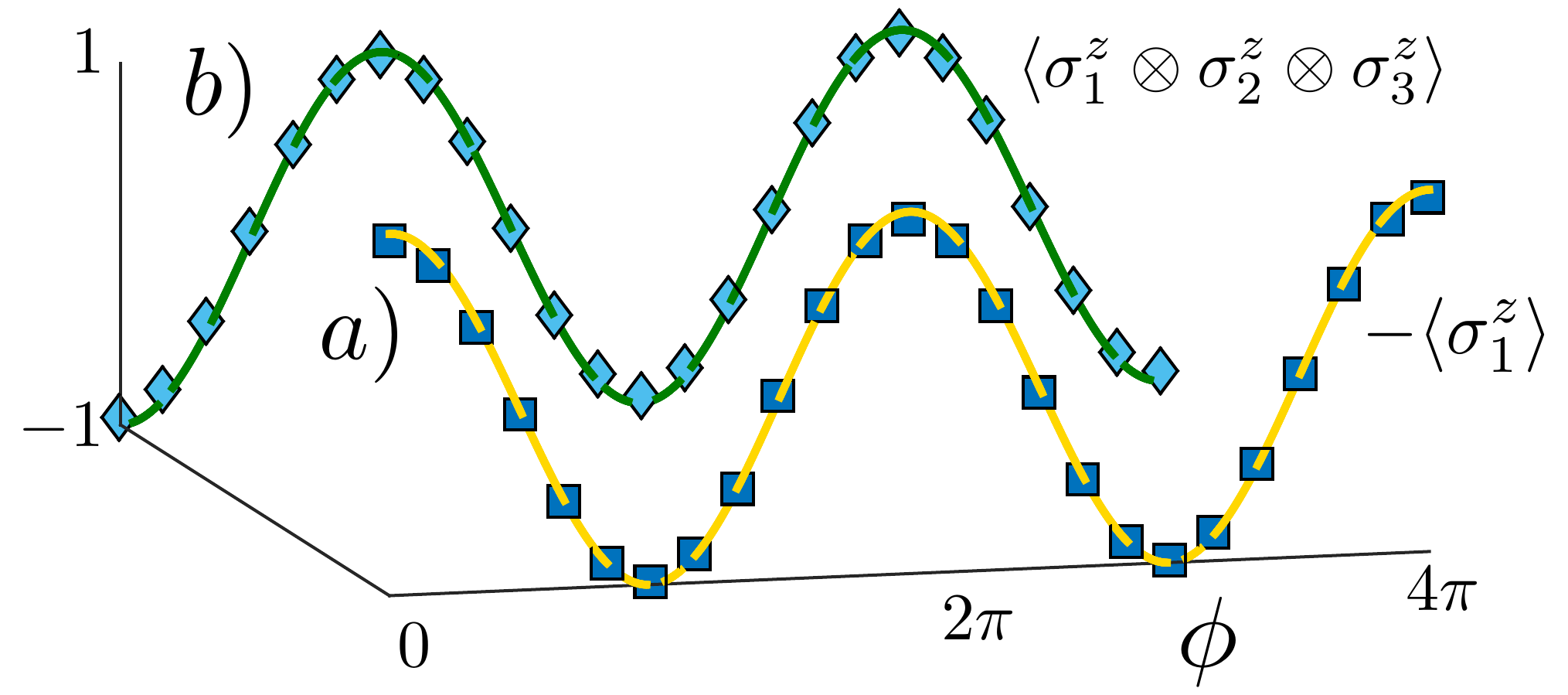}\caption{\label{expectation} Expectation value of  a) the $\sigma^z_1$ operator. Solid line, ideal result  according to the propagator in the second line of Eq.~(\ref{numerics}) while the squares correspond to the result when the sequence of gates in the first line of  Eq.~(\ref{numerics}) is applied. b) Expectation value of the delocalised  operator $\sigma^z_1\otimes \sigma^z_2 \otimes \sigma^z_3$. The solid line represents the ideal evolution while diamonds are the result of applying our method. All the points in the plot (squares and diamonds) correspond to the application of 3202 imperfect microwave pulses with a $\pi$ pulse time of $12.5$ ns, and a total evolution  time of $\approx 0.7$ ms to generate the propagator in Eq.~(\ref{numerics}). Note that this value is independent of the achieved phase $\phi$.}
\end{figure}

\section{Numerical results} 
We have numerically simulated the three-body nuclear time evolution operator
$\exp{(i 2^3 \phi \ I^1_x I^2_x I^3_x)}$, that gives rise to maximally entangled GHZ-like states \cite{Greenberger99} when the nuclear register is initialized into  the  state $|\downarrow \downarrow \downarrow~\rangle$. The latter can be prepared by polarisation transfer from the electron spin, see for example~\cite{Taminiau14, Waldherr2014, Cramer16}, or by dynamical nuclear polarisation (DNP)~\cite{London13, Chen15, Scheuer16}. The three-body nuclear propagator can be achieved by  applying the sequence in Eq.~(\ref{recipe}) with $Q_{N}=\prod_{n=1}^{3}Q_{j_{n}}^{\alpha_n}(\varphi_n)$ where  $\alpha_1=\alpha_2 = \alpha_3 =x$ and $\varphi_1=\varphi_2=\varphi_3=\frac{\pi}{2}$. More specifically, this is
\begin{eqnarray}\label{numerics}
U_\phi &=& Q_{1}^x\bigg(\frac{\pi}{2}\bigg) Q_{2}^x\bigg(\frac{\pi}{2}\bigg) Q_{3}^x\bigg(\frac{\pi}{2}\bigg) X_{2\phi+\pi} Q_{1}^x\bigg(\frac{\pi}{2}\bigg) Q_{2}^x\bigg(\frac{\pi}{2}\bigg) Q_{3}^x\bigg(\frac{\pi}{2}\bigg) X_{\pi}\nonumber \\
&=& - \exp{\bigg( i2^3 \phi \ \sigma_y I_1^x  I^x_2 I^x_3 \bigg)} = - \exp{\bigg( i\phi \ \sigma_y \sigma_1^x \sigma_2^x \sigma_3^x\bigg)}.
\end{eqnarray}

We selected a three qubit nuclear register such that $\vec{A}_1 = (2\pi)\times(-56, -32, -45)$ kHz, $\vec{A}_2 = (2\pi)\times(-7.6, 39, 52)$ kHz, $\vec{A}_3 = (2\pi)\times(-22, 13, 96)$ kHz, all of them corresponding to nuclei located in available positions of the diamond lattice, and the static magnetic field is $B_z = 0.65$ T, for more details see Appendix. Furthermore, and although not included in our theoretical description in Eqs.~(\ref{modelNV}) and (\ref{modelNV2}), we have also taken into account in the simulations the effect of  internuclear interactions. These produce a coupling between the $i$-th  and $j$-th nuclei of the form $g_{i,j}=\frac{\hbar \mu_0 \gamma^2_{^{13}\rm C}}{2 d_{i,j}^3} [1 - 3 (n_{i,j}^z)^2]$ with $d_{i,j}$ the relative distance between nuclei and $n_{i,j}^z$ the amplitude of the projection in $\hat{z}$ on their relative positioning vectors. In our case we have $g_{1,2} \approx -(2\pi) \times 20$ Hz, $g_{1,3}\approx -(2\pi) \times 10$ Hz, and $g_{2,3}\approx (2\pi) \times 7.5$ Hz. Figure~\ref{expectation} shows the evolution of the expectation value of a single nucleus $\sigma_1^z$ and of the delocalised operator $\sigma^z_1\otimes \sigma^z_2 \otimes \sigma^z_3$. The solid line corresponds to the ideal behavior while  squares and diamonds represent the result when our method is applied. Furthermore, we computed that the fidelity for the creation of a three-qubit GHZ-like nuclear state of the form $|\Psi\rangle = \exp{(i\frac{\pi}{4} \sigma_y \sigma_1^x \sigma_2^x \sigma_3^x  )} |y_{+}\rangle | \downarrow \downarrow \downarrow \rangle =  |y_{+}\rangle \frac{1}{\sqrt{2}}( | \downarrow \downarrow \downarrow \rangle + i   | \uparrow \uparrow \uparrow \rangle)$ is $ 98.8 \%$. This state has been prepared employing the same number of imperfect pulses, 3202,  than the one used for Fig~\ref{expectation}. In addition, in Appendix B  (Numerical simulations) we have included a plot that presents the impact of pulse-phase inaccuracy in our method.

\section{Further applications} 

The generation of these kind of gates, single- and $N$-qubit, allows to deal with problems involving fermionic interactions. It is known that any creation or annihilation  fermionic operator admits a form in terms of tensorial products of Pauli matrices when  the Jordan-Wigner transformation is applied~\cite{JW}. Hence, an appropriate application of our techniques would be of benefit to implement dynamics that include interacting fermions, e.g. quantum chemistry problems, in a solid-state quantum platform. Furthermore, the access to arbitrary multi-qubit spin propagators is of interest for simulating spin models with topological order~\cite{Nayak08}, as well as to generate dynamics and to perform measurements in different models of quantum computing as the case of deterministic quantum computation with one quantum bit, the DQC1 protocol,  that do not require to initially  polarise the nuclear register~\cite{Parker02, Boixo08}.

\section{Conclusions} 

We presented a protocol that allows  the generation of single and $N$-qubit quantum gates between nuclear spins in a solid state register such as diamond, as well as to measure highly delocalised nuclear spin correlators.
These gates are mediated by an effective electron spin, for example a NV center, externally controlled with microwave radiation in a dynamical decoupling scheme to assure selective entangling gates and electron spin state protection. The method is general and, therefore, applicable to other lattice defects as silicon carbide or germanium vacancy centers.

\section{Acknowledgements} 

The authors acknowledge J. F. Haase, T. Theurer, and R. Puebla for their useful comments on the manuscript. This work was supported by the Alexander von Humboldt Foundation, the ERC Synergy grant BioQ, the EU projects DIADEMS, EQUAM and HYPERDIAMOND as well as the
DFG via the SFB TRR/21 and the SPP 1601. J. C. acknowledges Universität Ulm for a Forschungsbonus.

\section{Appendix}
\subsection{Electron-nuclei many body gate}
Here we show how to derive Eq.~(\ref{Uphi}).
\begin{eqnarray}
 U_{\phi} & = & Q_{j_{1}}^{\alpha_{1}}(\varphi_{1})Q_{j_{2}}^{\alpha_{2}}(\varphi_{2})X_{2\phi+\pi}Q_{j_{1}}^{\alpha_{1}}(\varphi_{1})Q_{j_{2}}^{\alpha_{2}}(\varphi_{2})X_{\pi}\nonumber\\
 &=& \bigg[\exp{(i\varphi_1\sigma_z I_{j_1}^{\alpha_1})}  \exp{(i\varphi_2\sigma_z I_{j_2}^{\alpha_2})} \bigg]  \exp{(i\phi\sigma_{x})} \nonumber\\
 &&\ i\sigma_x \bigg[\exp{(i\varphi_1\sigma_z I_{j_1}^{\alpha_1})}  \exp{(i\varphi_1\sigma_z I_{j_2}^{\alpha_2})} \bigg] \ i\sigma_x \nonumber\\
 &=& e^{i\pi} \exp{\big[i\phi \  e^{(i\varphi_1\sigma_z I_{j_1}^{\alpha_1})}   e^{(i\varphi_2\sigma_z I_{j_2}^{\alpha_2})} \ \sigma_{x} \ e^{(-i\varphi_1\sigma_z I_{j_1}^{\alpha_1})}   e^{(-i\varphi_2\sigma_z I_{j_2}^{\alpha_2})}}\big]  \nonumber \\
&=&e^{i\pi} \exp{\big[i\phi \ \sigma_{x} \ e^{(-i 2 \varphi_1\sigma_z I_{j_1}^{\alpha_1})}   e^{(-i 2 \varphi_2\sigma_z I_{j_2}^{\alpha_2})}}\big] \nonumber\\
&=& e^{i\pi} \exp\bigg\{i\phi \ \sigma_{x}  [\cos{(\varphi_1)} - 2i \sin{(\varphi_1)}\sigma_z I_{j_1}^{\alpha_1}  ] \nonumber \\ 
&\times& [\cos{(\varphi_2)} - 2i \sin{(\varphi_2)}\sigma_z I_{j_2}^{\alpha_2}  ] \bigg\}.
\end{eqnarray}
Now, if the global phase factor $e^{i\pi}$ is neglected, we get the result at Eq.~(\ref{Uphi}).
Note that we have used $Q_{j}^{\alpha}(\varphi)=\exp\left(i\varphi\sigma_{z}I_{j}^{\alpha}\right)$ and $X_{\phi}=e^{i\phi\frac{\sigma_{x}}{2}}$ in concordance with the definitions in the main text. 
\begin{figure}[t]
\hspace{-0.40 cm}\includegraphics[width=0.75\columnwidth]{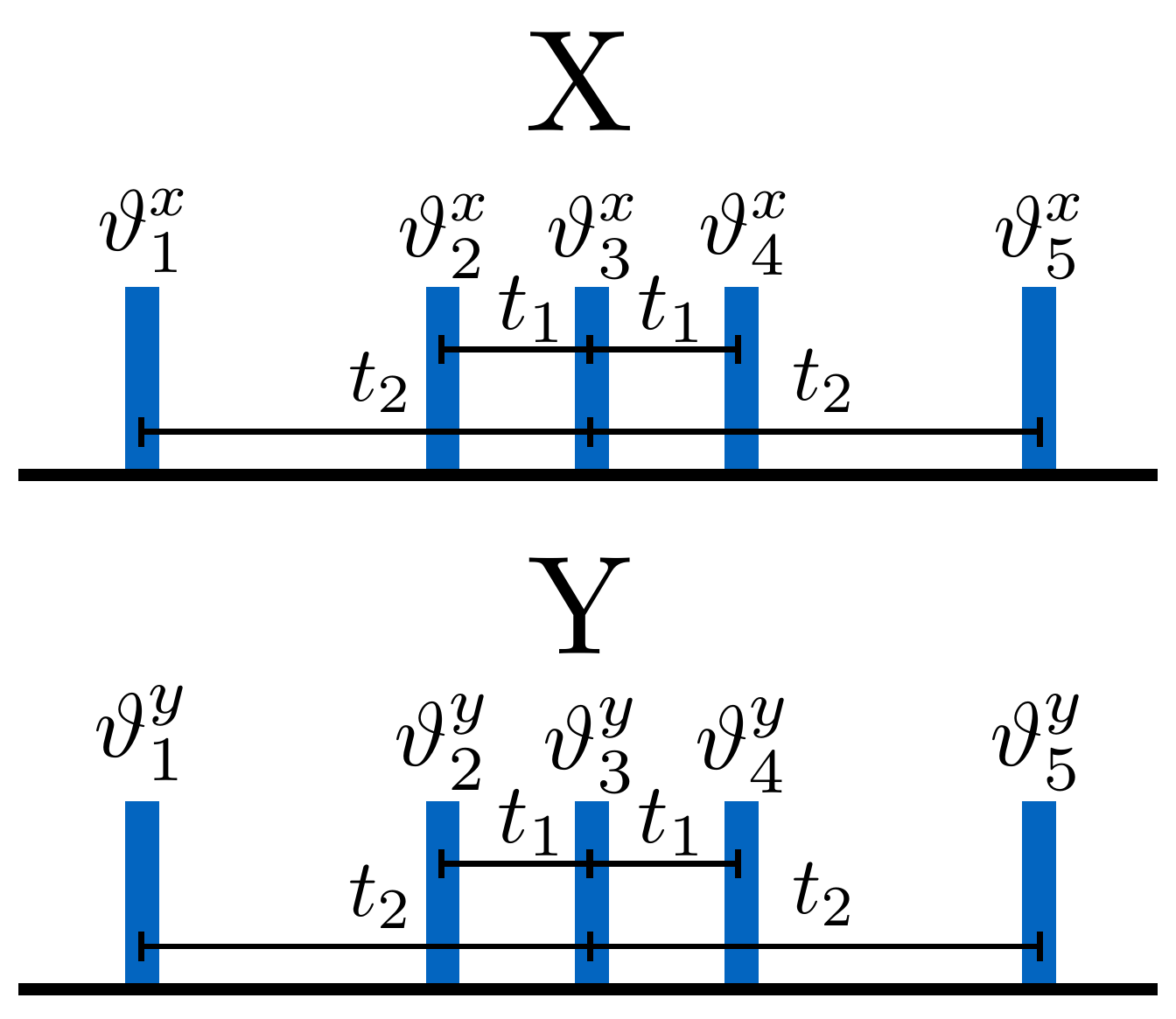}\caption{\label{figpulses} Internal structure of each X or Y composite pulses in terms of the 5 elementary $\pi$-pulses. These $\pi$-pulses are arranged symmetrically, see the tunable time distances $t_1$ and $t_2$  that distribute the pulses, with respect to the central $\pi$-pulse. }
\end{figure}
\subsection{Numerical simulations}
To obtain Fig.~[1], we chose a electron nuclear configuration such that the hyperfine vectors for the three $^{13}$C nuclei are 
\begin{eqnarray}
\vec{A}_1 &=& (2\pi)\times (-56, -32, -45) \ {\rm kHz}, \nonumber\\
\vec{A}_2 &=& (2\pi)\times (-7.6, 39,  52) \ {\rm kHz}, \nonumber\\
\vec{A}_3 &=& (2\pi)\times (-22 , 13,  96) \ {\rm kHz}.
\end{eqnarray}
The static $B_z$ field is aligned with the NV axis (the $z$ axis) and has a value of  $0.65$ T. We drive the electron spin with microwave pulses in the form of top-hat functions with a $\pi$-pulse time of $12.5$ ns. The microwave sequence is made of three different steps, one for each of the operations $Q_1^x(\frac{\pi}{2})$, $Q_2^x(\frac{\pi}{2})$ and $Q_3^x(\frac{\pi}{2})$, driven by an appropriate dynamical decoupling sequence. Note that each of these steps is repeated twice because the gates $Q_1^x(\frac{\pi}{2})$, $Q_2^x(\frac{\pi}{2})$ and $Q_3^x(\frac{\pi}{2})$ appear in front and behind the central gate $X_{2\phi+\pi}$ in the first line of Eq.~(15) in the main text. 

To implement each step, we use repetitively several  AXY-$8$ blocks~\cite{Casanova15} where each block has the following structure XYXYYXYX with  X (or Y) being a composite pulse containing 5 $\pi$-pulses, see Fig.~\ref{figpulses}. In addition one should note that each $\pi$-pulse is applied along an axis in the x-y plane determined by the phase $\vartheta_j^{x,y}$. This can be seen noting that each $\pi$-pulse is generated through the Hamiltonian $H_{\rm c} = \Omega (|1\rangle\langle0| e^{i\vartheta} + |0\rangle\langle1| e^{-i\vartheta})$. To assure robustness, see~\cite{Casanova15}, we set these phases as  $\vartheta_1^{x} =\pi/6$, $\vartheta_2^{x} =0$, $\vartheta_3^{x} =\pi/2$, $\vartheta_4^{x} =0$, and $\vartheta_5^{x} =\pi/6$, while the $\vartheta_j^{y}$ are shifted by an amount $\pi/2$ with respect to $\vartheta_j^{x}$. That is $\vartheta_j^{y} = \vartheta_j^{x} + \pi/2$
\begin{figure}[b]
\hspace{-0.40 cm}\includegraphics[width=0.75\columnwidth]{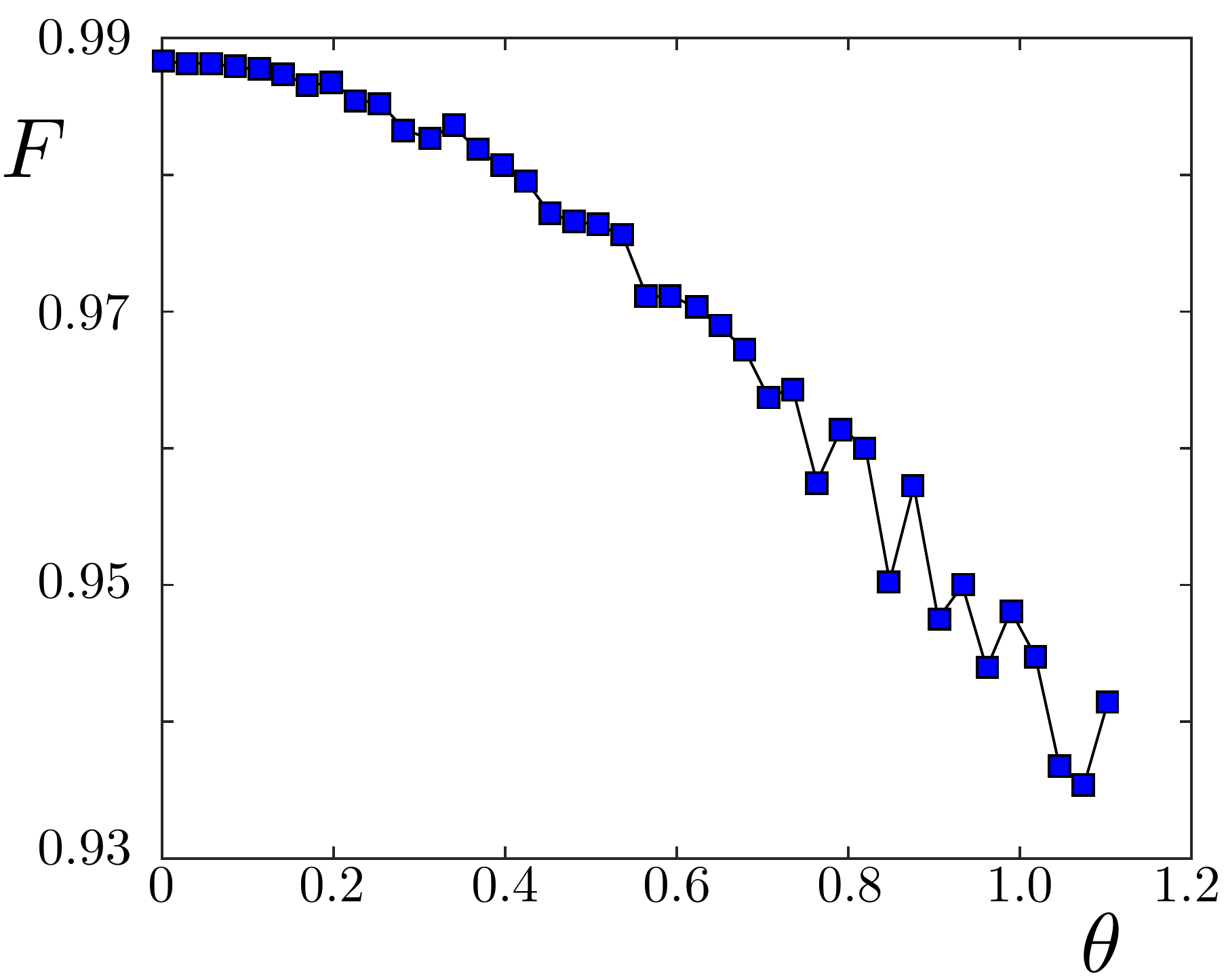}\caption{\label{Fdegrees} Fidelity of state preparation, $F$,  under the presence of a random error $\pm \theta$ (in degrees) in the pulse-phases $\vartheta_{j}^{x,y}$ see Fig.~\ref{figpulses}.  We observe a fidelity decrease for larger values of $\theta$. Each point in the plot has been taken by averaging 100 runs of the scheme in Eq.~(\ref{numerics}).}
\end{figure}
The gate $Q_1^x(\frac{\pi}{2})$ required $\approx 69\ \mu s$  to be displayed (we have used the 11-th harmonic of the decoupling sequence and 440 microwave pulses, i.e. 88 robust composite pulses. The other gates $Q_2^x(\frac{\pi}{2})$ and $Q_3^x(\frac{\pi}{2})$ are implemented in $\approx 107\ \mu s$ (440 microwave pulses, i.e. 88 robust composite pulses) and $\approx 177\ \mu s$ (720 microwave pulses, i.e. 144 robust composite pulses) respectively by making use of the 17-th harmonic in both cases. Each block has a distinct interpulse spacing to assure that the final achieved phase for each of the the $Q_j^x$ gates is $\pi/2$. In addition, one can calculate that the largest time to execute an AXY-8 block is $\approx 9.8 \ \mu$s, which corresponds to the case of the $Q_3^x(\frac{\pi}{2})$ gate. One can get this time interval by dividing the total time to implement $Q_3^x(\frac{\pi}{2})$, $177 \ \mu$s,  by the number of AXY-$8$ blocks that is equal to $144/8 = 18$. In the same manner, for the other gates it is possible to obtain that the times to display each individual AXY-$8$ block are $6.9\ \mu$s and $9.8 \ \mu$s. Hence, as these time intervals are very small with respect to the correlation time of the microwave's Rabi frequency fluctuation ($\approx$ 1 ms, see~\cite{Cai12}) we will consider this error as constant.

Finally, in Fig.~\ref{Fdegrees} we show the behaviour of the fidelity for a situation of growing pulse-phase errors. More specifically, we have simulated the state preparation fidelity of the same three-qubit GHZ state in the main text where each pulse-phase has a random error of $\pm \theta$ that accounts for the possible inaccuracy on the pulse-phase selection. Each point in the plot has been calculated by averaging the results of 100 runs of our gate scheme.

\end{document}